# Polymer translocation through a nanopore: A two-dimensional Monte Carlo Study

Kaifu Luo[1,*a], T. Ala-Nissila[1,2*b] and See-Chen Ying[2*c]

[1] Laboratory of Physics, Helsinki University of Technology, P.O. Box 1100, FIN-02015 HUT, Finland

[2] Department of Physics, Box 1843, Brown University, Providence, RI 02912-1843, U.S.A.

**ABSTRACT** We investigate the problem of polymer translocation through a nanopore in the absence of an external driving force. To this end, we use the two-dimensional (2D) fluctuating bond model with single-segment Monte Carlo moves. To overcome the entropic barrier without artificial restrictions, we consider a polymer which is initially placed in the middle of the pore, and study the escape time $\tau$ required for the polymer to completely exit the pore on either end. We find numerically that $\tau$ scales with the chain length $N$ as $\tau \sim N^{1+2\nu}$, where $\nu$ is the Flory exponent. This is the same scaling as predicted for the translocation time of a polymer which passes through the nanopore in one direction only. We examine the interplay between the pore length $L$ and the radius of gyration $R_g$. For $L \ll R_g$, we numerically verify that asymptotically $\tau \sim N^{1+2\nu}$. For $L \gg R_g$, we find $\tau \sim N$. In addition, we numerically find the scaling function describing crossover between short and long pores. We also show that $\tau$ has a minimum as a function of $L$ for longer chains when the radius of gyration along the pore direction, $R_\parallel \approx L$. Finally, we demonstrate that the stiffness of the polymer does not change the scaling behavior of translocation dynamics for single-segment dynamics.

---

a E-mail: luokaifu@yahoo.com
b E-mail: Tapio.Ala-Nissila@tkk.fi
c E-mail: ying@physics.brown.edu

## I. Introduction

The translocation of biopolymers through nanometer-scale pores is one of the most crucial processes in biology, such as DNA and RNA translocation across nuclear pores, protein transport through membrane channels, and virus injection.[1-3] Moreover, translocation processes might eventually prove useful in various technological applications, such as rapid DNA sequencing,[4-5] gene therapy and controlled drug delivery, *etc*.[6] In addition to its biological relevance, the translocation dynamics is also a challenging topic in polymer physics. Accordingly, the polymer translocation has attracted a considerable number of experimental,[7-14] theoretical[15-28] and numerical studies.[29-36]

The translocation of a polymer through a nanopore faces a large entropic barrier due to the loss of a great number of available configurations. In order to overcome the barrier and to speed up the translocation, an external field or interaction is often introduced. The possible driving mechanisms include an external electric field, a chemical potential difference, or selective adsorption on one side of the membrane. For example, in 1996, Kasianowicz *et al*.[7] reported that an electric field can drive single-stranded DNA and RNA molecules through the α-hemolysin channel of inside diameter 2 nm and that the passage of each molecule is signaled by the blockade in the channel current.

Inspired by the experiments,[7] a number of recent theories[15-28] have been developed for the dynamics of polymer translocation. Even without an external driving force, polymer translocation remains a challenging problem. To this end, Park and Sung[16] and Muthukumar[19] considered equilibrium entropy of the polymer as a function of the position of the polymer through the nanopore. The geometric restriction leads to an entropic barrier. Standard Kramer analysis of diffusion through this entropic barrier yields a scaling prediction of the translocation time $\tau_{tran} \sim N^2$ for long chains. However, as Chuang *et al.*[23] noted, this quadratic scaling behavior is at best only marginal for phantom polymers and cannot be correct for a self-avoiding polymer. The reason is that the equilibration time $\tau_{equil} \sim N^2$ for a phantom polymer and $\tau_{equil} \sim N^{1+2\nu}$ for a self-avoiding polymer, where $\nu$ is the Flory exponent ($\nu = 3/4$ and $3/5$ in 2D and 3D, respectively). Thus the exponent for $\tau_{equil}$ is larger than two for self-avoiding polymers, implying that the translocation time is shorter than the equilibration time of a long chain, thus rendering the concept of equilibrium entropy and the ensuing entropic barrier inappropriate for the study of translocation dynamics. Chuang *et al.*[23] performed numerical simulations with Rouse dynamics for a 2D lattice model to study the translocation for both phantom and self-avoiding polymers. They decoupled the translocation dynamics from the diffusion dynamics outside the pore by imposing the artificial restriction that the first monomer, which is initially placed in the pore, is never allowed to cross back out of the pore, see Fig.1(a).[23] We will refer to the translocation time obtained this way as $\tau_{tran}$. Their results show that for large *N*, translocation time $\tau_{tran}$ scales approximately in the same manner as equilibration time, but with a larger prefactor.

In the present work we consider a polymer which is initially placed symmetrically in the middle of the pore, as in Fig.1(b). In this case, without any external driving force or restriction, the polymer escapes from the hole either to the left or the right side of the pore in an average time defined as the escape time $\tau$.[37] It is clear that $\tau_{tran}$ and $\tau$ are different. Namely, the translocation time $\tau_{tran}$ includes events where the middle segment reaches the center of the pore but then the first segment returns to the entrance of the pore and the whole translocation process begins all over again. Numerically, $\tau$ can be sampled much more efficiently than $\tau_{tran}$, leading to a more accurate determination of the scaling behavior. We will show numerically that $\tau \sim N^{1+2\nu}$, in the same manner as found previously for $\tau_{tran}$. Recently, Wolterrink et al[28] have studied the translocation dynamics scaling for a 3D lattice model of a polymer. They have also found that $\tau$ scales as $\tau \sim N^{1+2\nu}$, in agreement with the present work.

In this study, we investigate the translocation dynamics in a 2D lattice model by focusing on $\tau$. In particular, we investigate the effect of varying the pore length on the polymer translocation. The dependence of the translocation dynamics on the stiffness of the polymer is also considered. The paper is organized as follows: In Sec. II we introduce the fluctuating bond model. In Sec. III using our approach we examine the polymer translocation through short and long pores. We obtain very accurate estimates for the scaling exponents as a function of $N$ and find the scaling function describing crossover between short and long pores. Our results are summarized in Sec. IV.

**II. The Fluctuating Bond Model**

The fluctuating bond (FB) model[38] combined with single-segment Monte Carlo (MC) moves has been shown to provide an efficient way to study many static and dynamic properties of polymers. Here we use the 2D lattice FB model for MC simulations of a self-avoiding polymer, where each segment excludes four nearest and next nearest neighbor sites on a square lattice. The bond lengths $b_l$ are allowed to vary in the range $2 \leq b_l \leq \sqrt{13}$ in units of the lattice constant, where the upper limit prevents bonds from crossing each other. The stiffness of the chain is controlled through an angle dependent potential[39,40] $\frac{U}{k_B T} = -\frac{J}{k_B T} \sum_{j=1}^{N_{FB}-1} \cos(\phi)$, where $J$ is the interaction strength, $N_{FB}$ is the number of segments in the chain, $\phi$ is the angle between two adjacent bonds, $k_B$ is the Boltzmann constant and $T$ is the absolute temperature. Dynamics is introduced in the model by Metropolis moves of a single segment, with a probability of acceptance min[$e^{-\Delta U/k_B T}$,1], where $\Delta U$ is the energy difference between the new and old states. As to an elementary MC move, we randomly select a monomer and attempt to move it onto an adjacent lattice site (in a randomly selected direction). If the new position does not violate the excluded-volume or maximal bond-length restrictions, the move is accepted or rejected according to Metropolis criterion. $N$ elementary moves define one MC time step.

**III. Results and Discussion**

**1. Polymer translocation through a short nanopore**

In this section, we present the results for the escape time $\tau$ for a lattice model of polymers. For the same model, we also studied the translocation time $\tau_{tran}$ defined by Chuang et al [23] using the restriction that the first polymer cannot back out of the pore. We used the same pore size for both cases (length of $L = 3$ and width of $w = 2$ lattice units). Numerical studies were done for a number of different chain lengths $N$, with several thousand runs for each case.

In Fig. 2(a) we show the average translocation time $\tau_{tran}$ and $\tau$ as a function of the polymer length $N$ for the case with $J = 0$. The log-log plot of Fig. 2(a) is shown in Fig. 2(b). When $N > 16$, the results vs. $N$ follows scaling to a good degree of accuracy. We find that $\tau_{tran} \sim N^{2.41\pm0.01}$. The scaling for the escape time is $\tau \sim N^{2.50\pm0.01}$. The result is very close to the expected value of $1 + 2\nu = 2.5$, and in agreement with the 3D numerical values obtained in Ref. 28.

We have also examined $\tau$ for a relatively stiff polymer, by setting $J/k_BT = 5$.[39] In Fig. 3 we show $\tau$ as a function of $N$. We find that $\tau \sim N^{2.58\pm0.01} \sim N^{1+2\nu}$, which confirms that the stiffness of the polymer does not affect the scaling behavior of $\tau$ for single-segment dynamics.

**2. Polymer translocation through a long pore**

Next, we consider the influence of the pore length $L$ on $\tau$. In the simulations, the width of the pore is chosen as $W=7$. For a successful passage through the pore of length $L$, the mass center of the polymer moves a distance of $L/2 + R_g$, so the time it takes can be estimated to be

$$\tau \sim \frac{(R_g + L/2)^2}{D} \sim \begin{cases} \frac{R_g^2}{D} \sim N^{1+2\nu} & L \ll R_g \\ \frac{L^2}{D} \sim N^1 & L \gg R_g \end{cases}. \quad (1)$$

This result indicates that there is a crossover as a function of $R_g/L$ such that for $L \ll R_g$,

scaling follows the previous short pore result, while for $L \gg R_g$, the scaling behavior changes to $\tau \sim N$. Fig. 4(a) shows our numerical data, where crossover is clearly seen. In Fig. 4(b) we show data for a very long pore with $L = 800$, which confirms the linear scaling behavior.

In general, we can write a scaling form for $\tau$ as

$$\tau \sim \frac{R_g^2}{D} f\left(\frac{R_g}{L}\right), \tag{2}$$

where $f(x)$ is a scaling function. Using Eq. (1), we obtain

$$\tau \frac{D}{R_g^2} \sim f\left(\frac{R_g}{L}\right) \sim \begin{cases} const. & L \ll R_g \\ \frac{L^2}{R_g^2} & L \gg R_g \end{cases}. \tag{3}$$

Using the data in Fig. 4, we plot the scaling function $f(x)$ in Figs. 5(a) and (b). From Fig. 5a, we find that $f(x) \sim const.$ for $L \gg R_g$, and in Fig. 5(b) we show the other limit on a log-log scale, confirming the predicted $x^{-2}$ behavior. We note that recently, Slonkina el al.[22] theoretically examined the polymer translocation through a long pore. Following the approach of Sung and Park[16] and Muthukumar[19], for $L > R_g$ $\tau_{tran} \sim (L/a - N)^2$ is obtained, where $a$ is the segment length. This result means that $\tau_{tran} \sim L^2$ for $L \gg R_g$, in contrast of our prediction here for $\tau \sim N$ in the large $L$ limit.

A particularly interesting question concerns the influence of the pore length to the actual translocation dynamics with fixed $N$. This problem has been theoretically addressed by Muthukumar[20] who investigated the free energy barrier and average translocation time for the movement of a single Gaussian chain from one sphere to another larger sphere through pores of different lengths. In Ref. 20, it was found that $\tau_{tran}(L)$ has a minimum for an "optimal" value of $L_0$, where translocation is fastest and a threshold value $L_c$. This result was explained to be due to interplay between polymer entropy and pore-polymer interaction energy. In the first regime where $L < L_c$, $\tau_{tran}(L)$ first decreases and then increase with $L$, the entopic barrier mechanism dominates polymer translocation. In the other regime $L > L_c$, $\tau_{tran}(L)$ increases with $L$, and polymer translocation is controlled by the pore-polymer interaction. The existence of $L_c$ thus corresponds to an apparent cancellation between the gain in translocation rate arising from the entropic part and the loss in the rate associated with the pore-polymer interaction.

In Fig.6 we show our numerical data for a fixed chain of length $N = 51$. Most strikingly, our result shows that the two regimes for the dependence of $\tau$ on the pore length are present here without any explicit pore-polymer interaction potential. Thus, the existence of an optimal pore length is a generic phenomenon in polymer translocation. We find that the "optimal" value of $L_0$, corresponds to the radius of gyration of the polymer along the pore direction, $R_\parallel$. The escape of the polymer consists of two steps. In the first step with average duration $\tau_1$, one end of the polymer reaches an edge of the pore. During the second step of average duration $\tau_2$, the other end of the polymer reaches this edge of the pore. For $L > L_0$, $\tau_1 = \tau_1(L)$ increases with $L$ because the polymer has to move a longer distance for increasing $L$, and $\tau_2 = \tau_2(N)$ depends on the chain length $N$, but is almost independent of the pore length. Thus the total time, $\tau = \tau_1(L) + \tau_2(N)$, increases with $L$. For $L < L_0$, $\tau$ goes down with increasing $L$. Further support for the existence of an optimal pore length comes from the results of Slonkina el al.[22] for an ideal polymer where a minimum translocation time occurs at $L_0 = Na$. Our simulation results show that $L_0 \approx R_\parallel$, but for narrow enough pore $R_\parallel \approx Na$.

**Conclusion**

In this paper, the polymer translocation through a nanopore in the absence of an external driving force is examined both theoretically and numerically. To overcome the entropic barrier, we consider the translocation dynamics of a polymer that is initially placed in the middle of the pore in a symmetric position, instead of using a restriction that the first monomer is never allowed to crossing back out of the pore. Our numerical results show that accurate estimates for the scaling exponents of the escape time as a function of $N$ are obtained. Our theory predicts that the length of the pore plays a very important role in polymer translocation dynamics. For $L \ll R_g$, the escape time $\tau$ with polymer length $N$ satisfies $\tau \sim N^{1+2\nu}$, while for $L \gg R_g$, $\tau \sim N$ is observed. We also numerically find the scaling function describing crossover between short and long pores and show that $\tau$ has a minimum as a function of $L$ for longer chains. In addition, our numerical results show that the stiffness of a polymer does not change the scaling behavior of translocation dynamics.

**Acknowledgement** This work has been supported in part by a Center of Excellence grant from the Academy of Finland.

**References**

1) B. Alberts and D. Bray, *Molecular Biology of the Cell* (Garland, New York, 1994).

2) J. Darnell, H. Lodish, and D. Baltimore, *Molecular Cell Biology* (Scientific American Books, New York, 1995).

3) R. V. Miller, Sci. Amer. **278**, 66 (1998).

4) J. Han, S. W. Turner, and H. G. Craighead, Phys. Rev. Lett. **83**, 1688 (1999).

5) S. W. P. Turner, M. Calodi, and H. G. Craighead, Phys. Rev. Lett. **88**, 128103 (2002).

6) D.-C. Chang, *Guide to Electroporation and Electrofusion* (Academic, New York, 1992)

7) J. J. Kasianowicz, E. Brandin, D. Branton, and D. W. Deaner, Proc. Natl. Acad. Sci. U.S.A. **93**, 13770 (1996).

8) M. Aktson, D. Branton, J. J. Kasianowicz, E. Brandin, and D. W. Deaner, Biophys. J. **77**, 3227 (1999).

9) A. Meller, L. Nivon, E. Brandin, J. A. Golovchenko, and D. Branton, Proc. Natl. Acad. Sci. U.S.A. **97**, 1079 (2000).

10) S. E. Henrickson, M. Misakian, B. Robertson, and J. J. Kasianowicz, Phys. Rev. Lett. **85**, 3057 (2000).

11) A. Meller, L. Nivon, and D. Branton, Phys. Rev. Lett. **86**, 3435 (2001).

12) A. F. Sauer-Budge, J. A. Nyamwanda, D. K. Lubensky, and D. Branton, Phys.

Rev. Lett. **90**, 238101 (2003).

13) A. Meller, J. Phys.: Condens. Matter **15**, R581 (2003).

14) A. J. Storm, C. Storm, J. Chen, H. Zandbergen, J.-F. Joanny, C. Dekker, Nano Lett., **5**, 1193 (2005).

15) S. M. Simon, C. S. Reskin, and G. F. Oster, Proc. Natl. Acad. Sci. U.S.A. **89**, 3770 (1992).

16) W. Sung and P. J. Park, Phys. Rev. Lett. **77**, 783 (1996)

17) P. J. Park and W. Sung, J. Chem. Phys. **108**, 3013 (1998).

18) E. A. diMarzio and A. L. Mandell, J. Chem. Phys. **107**, 5510 (1997).

19) M. Muthukumar, J. Chem. Phys. **111**, 10371 (1999).

20) M. Muthukumar, J. Chem. Phys. **118**, 5174 (2003).

21) D. K. Lubensky and D. R. Nelson, Biophys. J. **77**, 1824 (1999).

22) E. Slonkina and A. B. Kolomeisky, J. Chem. Phys. **118**, 7112 (2003)

23) J. Chuang, Y. Kantor, and M. Kardar, Phys. Rev. E **65**, 011802 (2002).

24) Y. Kantor and M. Kardar, Phys. Rev. E **69**, 021806 (2004)

25) T. Ambjörnsson, S. P. Apell, Z. Konkoli, E. A. DiMarzio, and J. J. Kasianowicz, J. Chem. Phys. **117**, 4063 (2002).

26) U. Gerland, R. Bundschuh, and T. Hwa, Phys. Biol. **1**, 19 (2004).

27) A. Baumgärtner and J. Skolnick, Phys. Rev. Lett. **74**, 2142 (1995).

28) J. K. Wolterink, G. T. Barkema, and D. Panja,

http://arxiv.org/pdf/cond-mat/0509577.

29) S.-S. Chern, A. E. Cardenas, and R. D. Coalson, J. Chem. Phys. **115**, 7772 (2001).

30) H. C. Loebl, R. Randel, S. P. Goodwin, and C. C. Matthai, Phys. Rev. E **67**, 041913 (2003).

31) R. Randel, H. C. Loebl, and C. C. Matthai, Macromol. Theory Simul. **13**, 387 (2004).

32) Y. Lansac, P. K. Maiti, and M. A. Glaser, Polymer **45**, 3099 (2004)

33) C. Y. Kong and M. Muthukumar, Electrophoresis **23**, 2697 (2002)

34) Z. Farkas, I. Derenyi, and T. Vicsek, J. Phys.: Condens. Matter **15**, S1767 (2003).

35) P. Tian and G. D. Smith, J. Chem. Phys. **119**, 11475 (2003).

36) R. Zandi, D. Reguera, J. Rudnick, and W. M. Gelbart, Proc. Natl. Acad. Sci. U.S.A. **100**, 8649 (2003).

37) We note that in Ref. (28), the escape time discussed here is defined as the 'unthreading' time for a polymer in the pore.

38) I. Carmesin and K. Kremer, Macromolecules **21**, 2819 (1988)

39) T. Ala-Nissila, S. Herminghaus, T. Hjelt, and P. Leiderer, Phys. Rev. Lett. **76**, 4003 (1996).

40) T. Hjelt and I. Vattulainen, J. Chem. Phys. **112**, 4731 (2000).

**Figure captions**

**Fig.1** (a) A polymer is initially placed on the one side of the wall with the first monomer in the pore. (b) The middle of a polymer is initially placed in the center of the pore. The length and width of the pore are $L$ and $W$, respectively.

**Fig.2** (a) Average translocation time and escape time $\tau$ as a function of the chain length *N*. (b) A log-log plot of (a). The length and width of the pore are 3 and 2, respectively.

**Fig.3** Average $\tau$ as a function of the chain length of stiff polymers. The length and width of the pore are 3 and 2, respectively.

**Fig.4** (a) Average $\tau$ as a function of the chain length for polymer translocation through the long pore. (b) Average $\tau$ as a function of the polymer length for short polymer translocation through a very long pore. The pore width is 7.

**Fig.5** (a) Scaling function for polymer escape through a long pore. (b) A log-log plot of (a).

**Fig.6** (a) The effect of the pore length *L* on average $\tau$ for chain of length *N*=51. The pore width is 7.

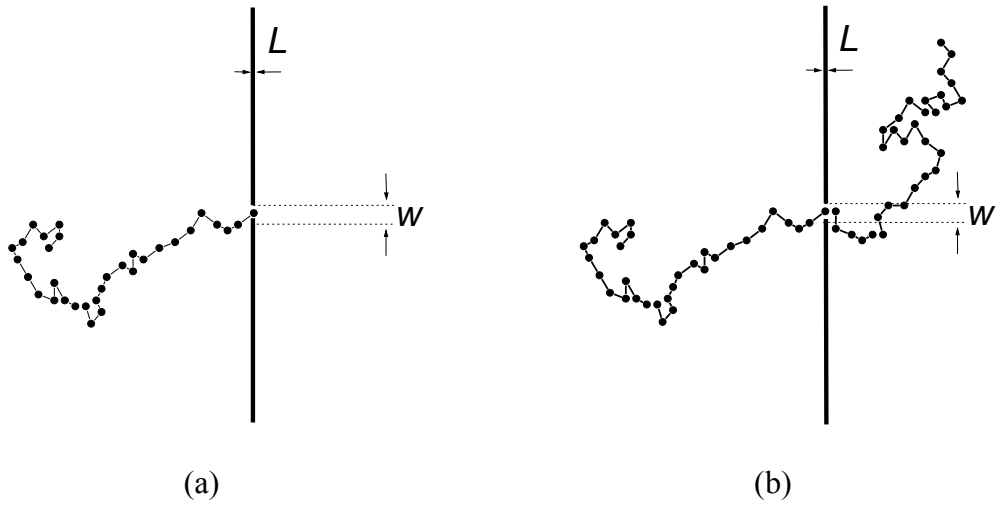

(a)                      (b)

**Fig.1**

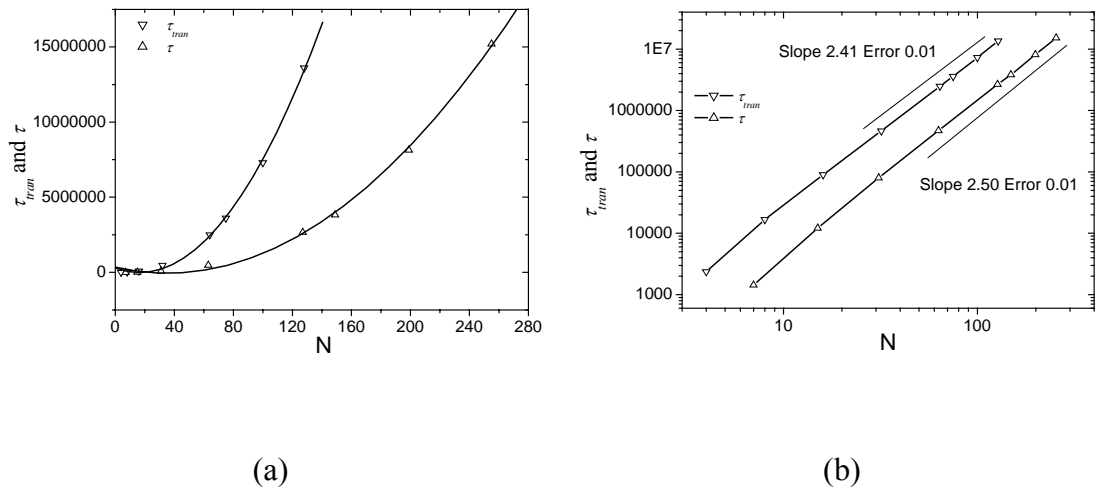

(a)                      (b)

**Fig.2**

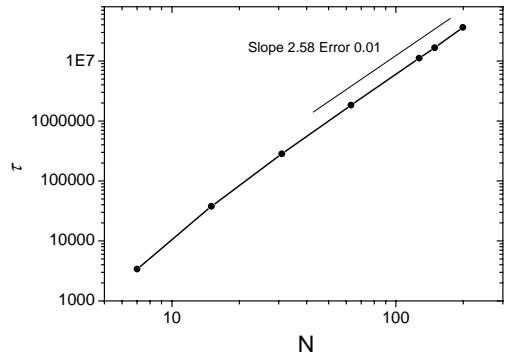

**Fig.3**

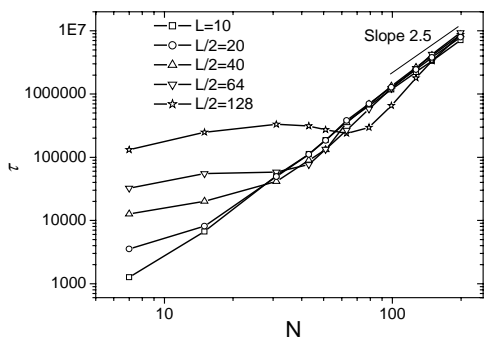  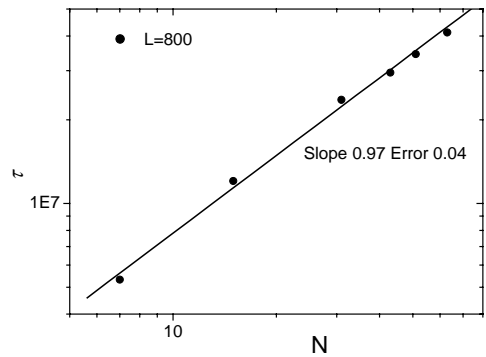

(a)                                  (b)

**Fig.4**

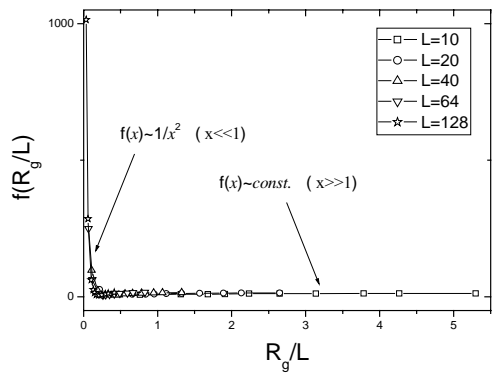 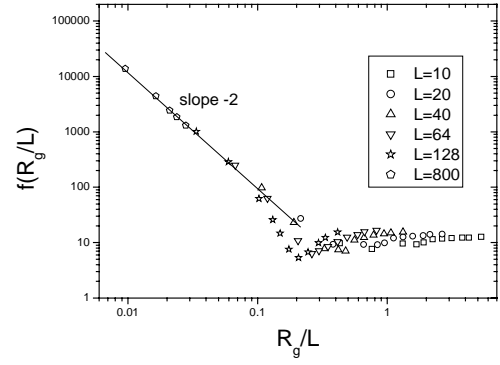

(a) (b)

**Fig.5**

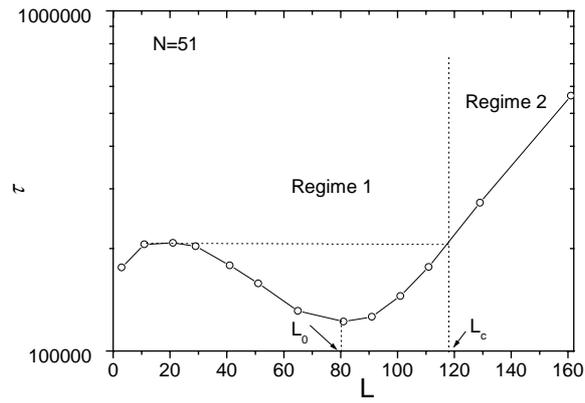

**Fig.6**